\documentclass[final]{cvpr}
\pdfoutput=1
\usepackage{times}
\usepackage{epsfig}
\usepackage{graphicx}
\usepackage{subfigure}
\usepackage{amsmath}
\usepackage{amssymb}

\usepackage[pagebackref=true,breaklinks=true,colorlinks,bookmarks=false]{hyperref}

\def\cvprPaperID{4964} 
\def\confYear{CVPR 2021}

\begin{document}
\title{Blocked and Hierarchical Disentangled Representation From Information Theory Perspective} 

\author{Ziwen Liu\\
University of Chinese Academy of Sciences\\
{\tt\small liuziwen18@mails.ucas.ac.cn}
\and
Mingqiang Li\\
Information Science Academy of China Electronics Technology Group Corporation\\
{\tt\small limingqiang14@mails.ucas.ac.cn}
\and
Congying Han\\
University of Chinese Academy of Sciences\\
{\tt\small hancy@ucas.ac.cn}
}

\maketitle
\begin{abstract}
   We propose a novel and theoretical model, blocked and hierarchical variational autoencoder (BHiVAE), to get better-disentangled representation. It is well known that information theory has an excellent explanatory meaning for the network, so we start to solve the disentanglement problem from the perspective of information theory. BHiVAE mainly comes from the information bottleneck theory and information maximization principle. Our main idea is that (1) Neurons block not only one neuron node is used to represent attribute, which can contain enough information; (2) Create a hierarchical structure with different attributes on different layers, so that we can segment the information within each layer to ensure that the final representation is disentangled. Furthermore, we present supervised and unsupervised BHiVAE, respectively, where the difference is mainly reflected in the separation of information between different blocks. In supervised BHiVAE, we utilize the label information as the standard to separate blocks. In unsupervised BHiVAE, without extra information, we use the Total Correlation (TC) measure to achieve independence, and we design a new prior distribution of the latent space to guide the representation learning. It also exhibits excellent disentanglement results in experiments and superior classification accuracy in representation learning.
\end{abstract}

\section{Introduction}
\paragraph{Disentanglement Representation}Learning an interpretable and disentangled representation of data to reflect the semantic meaning is what machine learning always pursues \cite{bengio2013representation,BengioLecun2007scaling,chen2016infogan,journals/neco/Schmidhuber92c}. Disentangled representation is defined in \cite{bengio2013representation} as:\emph{ a representation where a change in one dimension corresponds to a change in one factor of variation, while being relatively invariant to changes in other factors.} As far as our understanding is concerned, the fact that different dimensions do not affect each other means probabilistically independent.

As popular generative models, Variational Autoencoder (VAE) \cite{kingma2013auto} and Generative Adversarial Networks(GAN) \cite{goodfellow2014generative} have been applied in disentanglement. For example, InfoGAN \cite{chen2016infogan}, based on the GAN model, maximizes the mutual information between the small subset of the latent variables and the observations which makes the latent variables contain more information about the real data, hence increases the interpretability of the latent representation. Based on InfoGAN, FineGAN \cite{li2020mixnmatch,singh2019finegan} creates a hierarchical architecture that assigns the background, object shape, and object appearance to different hierarchy to generate images of fine-grained object categories. And VAE model, derived from autoencoder \cite{ackley1985learning} is also widely applied to representation learning, VAEs have been demonstrated their unique power to constrain representations disentanglement. For example, $\beta$-VAE \cite{Higgins2017betaVAELB}, $\beta$-TCVAE \cite{chen2018isolating}, FactorVAE \cite{pmlr-v80-kim18b} and so on \cite{ding2020guided} are able to get more disentangled representation.
\paragraph{Information Theory}
Information Theory has been proposed by Shannon in 1948 \cite{shannon1948mathematical}, which came from communication research. Mutual information is the fundamental metric for measuring the relationship about information between random variables. In representation learning, it has been applied widely \cite{pmlr-v80-belghazi18a,chen2016infogan,hjelm2019learning,Oord2018RepresentationLW}, with graph network \cite{ren2019heterogeneous,velickovic2019deep}, and gets some explanatory meaning on machine learning \cite{shwartz2017opening}. We can conclude the application as two ideas: The first one is Information Maximization Principle(InfoMax) \cite{bell1995information,linsker1988self}, which enforces representation to preserve more information about the input data through the transformers (CNN, GNN); some works \cite{chen2016infogan,hjelm2019learning,conf/iclr/VelickovicFHLBH19} regularize their original model with InfoMax term to get more informative and interpretable model. The other one is the Information Bottleneck(IB) theory \cite{shwartz2017opening,tishby99information,tishby2015deep}. It analyzes the process of information transmission and the loss through the networks.
IB theory considers the network process as a Markov chain and uses the Data Processing Inequality (DPI) \cite{cover1991elements} to explain the variation of information in deep networks. In 2015, Variational Information Bottleneck (VIB) method \cite{vib} offers a variational form of supervised IB theory.
Also, IB theory has been revealed a unique ability \cite{voloshynovskiy2019information} to explain how and why VAEs models design this architecture.
With this knowledge of disentanglement and information, we initiate our model, blocked and hierarchical variational autoencoder (BHiVAE), completely from information theory perspective to get better interpretability and controllability. In BHiVAE, because of the neural network's different ability to extract features with different net depth, we locate data factors into different layers. Furthermore, the weak expressiveness of single-neuron pushes us to use neuron blocks to represent features. We also discuss the supervised and unsupervised version model. In the supervised model, we utilize the label to separate the representation from feature information. In the unsupervised model, we give out a unique prior distribution to better meet our model and use additional discriminators to split information. Of course we give enough experiments in MNIST \cite{lecun1998mnist}, CelebA \cite{liu2015deep} and dSprite \cite{dsprites17} datasets to show the great performance in disentanglement. In summary, our work mainly makes the following contributions:
\begin{itemize}
	\item We approach the disentanglement problem for the first time entirely from an information theory perspective. Most previous works on disentanglement have been based on existing models and modified to fit the framework for solving entanglement problems.
	\vspace{-1mm}
	\item We present Blocked and Hierarchical Variational Autoencoder (BHiVAE) in both supervised and unsupervised cases. In the supervised case, we utilize the known feature information to guide the representation learning in each hierarchy; in the unsupervised case, we propose a novel distribution-based method to meet our neural block set.
    \item We perform experiments thoroughly on several public datasets, MNIST, dSprites and CelebA, comparing with VAE, $\beta$-VAE, FactorVAE, $\beta$-TCVAE, and Guided-VAE in several classic metrics. From the results, our method BHiVAE shows an excellent performance considering all the indicators together.
\end{itemize}
\section{Related Work}
In order to get disentangled representation, some previous work has made a significant contribution to it. Based on VAE, $\beta$-VAE \cite{Higgins2017betaVAELB} adds a coefficient weight to the KL-divergence term of the VAE loss and get a more disentangled representation. Mostly there is a significant advantage in that it trains more stably than InfoGAN. However, $\beta$-VAE sacrifices the reconstruction result at the same time. $\beta$-TCVAE \cite{chen2018isolating} and FactorVAE \cite{pmlr-v80-kim18b} explored this issue in more detail and found TC term is the immediate causes to promote disentanglement.

Guided VAE \cite{ding2020guided} also gives out a model using different strategies in supervised and unsupervised situations to get disentanglement representation. It uses additional discriminator to guide the representation learning and learn the knowledge about latent geometric transformation and principal components. This idea of using different methods with different supervised information inspires us. FineGAN \cite{singh2019finegan} based on InfoGAN, generates the background, object shape, and object appearance images respectively in different hierarchies, then combines these three images into true image. In FineGAN, what helps the disentanglement is the mutual information between the latent codes and each factor. And MixNMatch \cite{li2020mixnmatch}, developed from FineGAN, becomes a conditional generative model that learns disentangled representation and encodes different features from real image and then uses additional discriminators to match the representation to the prior distribution given by FineGAN model.

Previous works have made simple corrections to $\beta$-VAE or GAN model, adding some useful terms for solving disentanglement. In our work, we fully consider the disentanglement problem from information theory and then establish the BHiVAE model. Information theory and optimal coding theory \cite{cover1991elements, voloshynovskiy2019information} have shown that longer code can express more information. So in our model, instead of using only one dimension node to represent a ground-truth factor as in previous work, we choose multiple neural nodes to do so.

In the meantime, different ground-truth factors of data contain different levels of information, and the depth of the neural network affects the depth of information extracted, so a hierarchical architecture is used in our model for extracting different factor features at different layers. Therefore, in order to satisfy the requirement of disentanglement representation, i.e., the irrelevance between representation neural blocks, We only need to minimize the mutual information between blocks of the same layer due to characteristics of hierarchical architecture.

\begin{figure*}
	\centering
	\subfigure[Encoder part]{
		\label{encoder}
		\includegraphics[width=0.6\linewidth]{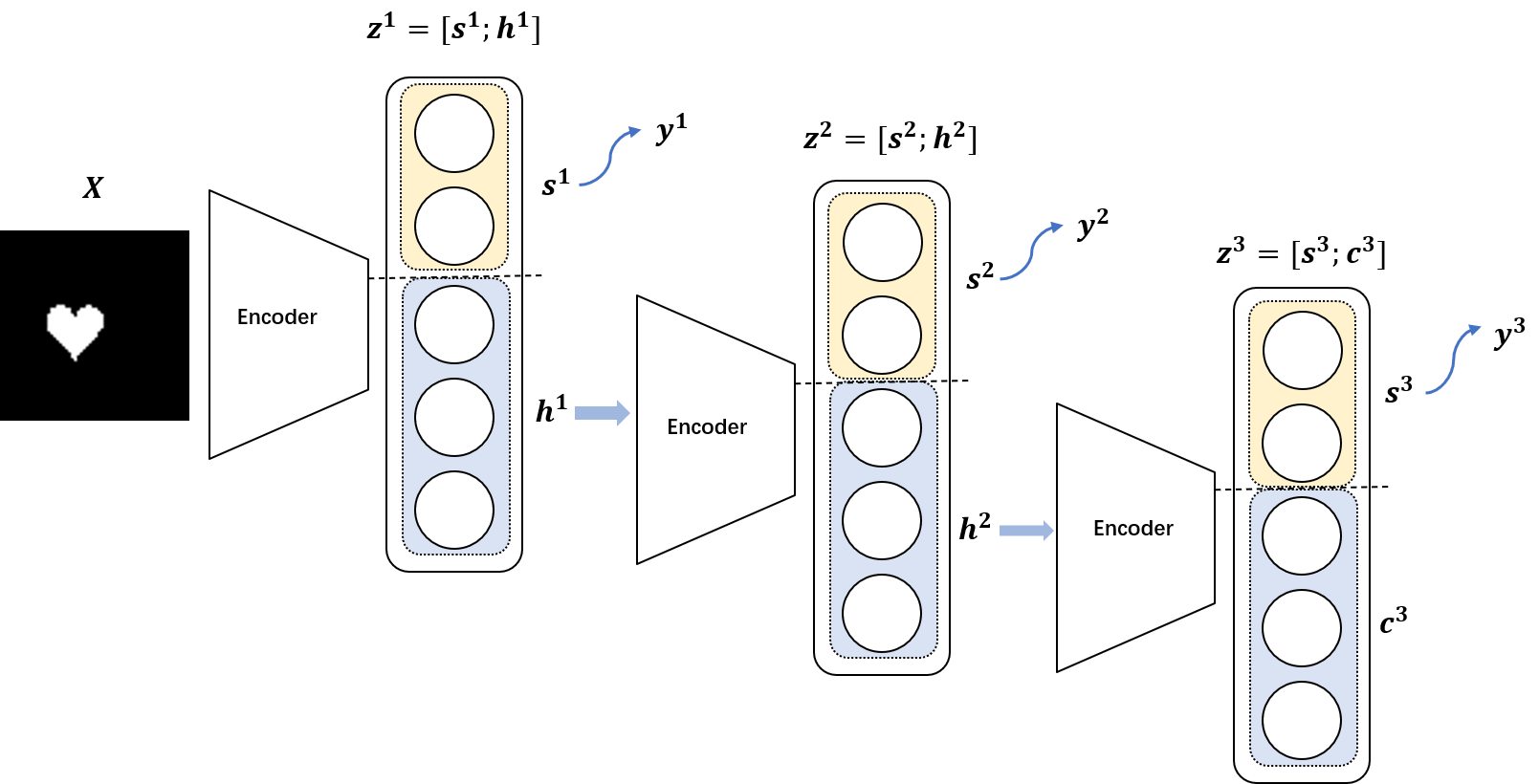}}
	\subfigure[Decoder part]{
		\label{decoder}
		\includegraphics[width=0.3\linewidth]{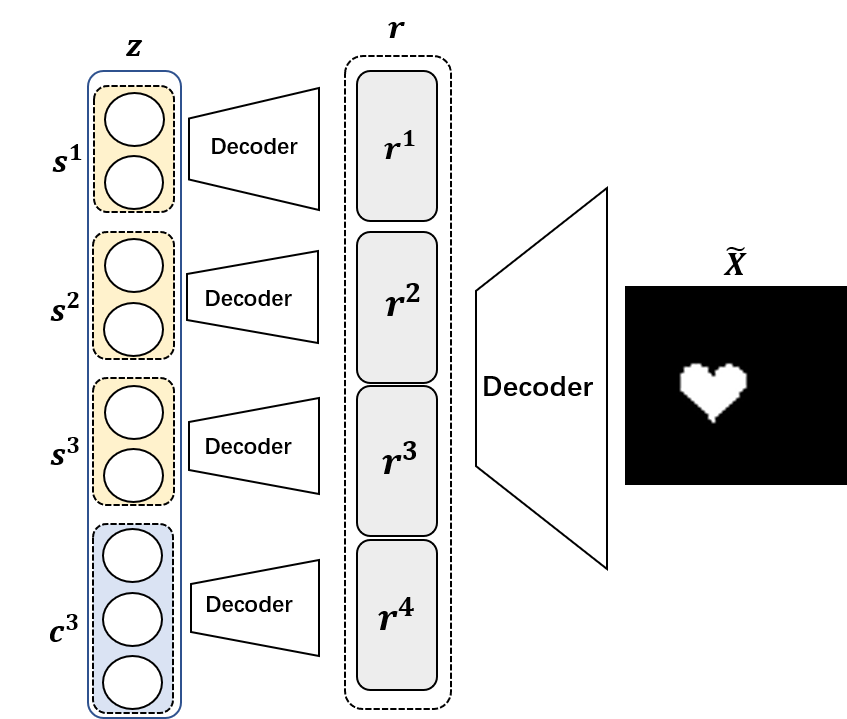}}
		\caption{Architecture of Hierarchical VAE model: Encoder part in the left-side and decoder in the right-side.}
	\label{fig:BHiVAE}
\end{figure*}
\vspace{-2mm}
\section{Proposed Method}

We propose our model motivated by IB theory and VAEs, like $\beta$-VAE, Factor-VAE, $\beta$-TCVAE, Guided-VAE, and FineGAN. Therefore, in this section, we first introduce the IB theory and VAEs models, and then we present our detailed model architecture and discuss supervised and unsupervised BHiVAE methods.
\subsection{Information Theory and VAEs}
IB theory aims to learn a representation $Z$ that maximizes the compression of informaiton in real data $X$ while maximizing the expression of target $Y$. So we can describe it as:
\begin{align}
	\min I(X;Z)-\beta I(Z;Y)
	\label{ib}
\end{align}
the target $Y$ is the attribute information under supervision, and is equal to $X$ under unsupervision \cite{voloshynovskiy2019information}.

In the case of supervised IB theory \cite{vib}, we can get the upper bound:
\begin{align}
	I_{\phi}(X;Z)-\beta I_{\theta}(Z;Y)\le& \mathbb{E}_{p_D(x)}[D_{KL}(q_{\phi}(z|x)\|p(z))]\nonumber\\
	&-\beta \mathbb{E}_{p(x,y)}[q_{\phi}(z|x)\log p_{\theta}(y|z)]
\end{align}
The first term represents the KL divergence between the posterior $q_{\phi}(z|x)$ and the prior distribution $p(z)$; and absolutely, the second term equals cross-entropy loss of label prediction.

And in the case of unsupevised IB theory, the we can rewrite the objective Eq. (\ref{ib}) as:
\begin{align}
	\min I_{\phi}(X;Z)-\beta I_{\theta}(Z;X)
	\label{uns_ib}
\end{align}
Unsupervised IB theory seems like generalization of VAEs model, with an encoder to learn representation and a decoder to reconstruct. $\beta$-VAE \cite{Higgins2017betaVAELB} is actually the upper bound of it:
\begin{align}
	\mathcal{L}_{\beta- VAE}=&\mathbb{E}_{p(x)}[D_{KL}(q_{\phi}(z|x)\|p(z))\nonumber\\
	&-\beta\mathbb{E}_{q_{\phi}(z|x)}[\log(p_{\theta}(x|z))]]
	\label{betaVAE}
\end{align}
FactorVAE \cite{pmlr-v80-kim18b} and $\beta$-TCVAE \cite{chen2018isolating} just add more weight on the TC term $\mathbb{E}_{q(z)}[\log\frac{q(z)}{\tilde{q}(z)}]$, which express the dependence across dimensions of variable in information theory, where $\tilde{q}(z)=\prod_{i=1}^nq(z_i)$.

We build our BHiVAE model upon above works and models. We focus on information transmission and loss through the whole network, and then achieve it through different methods.

\subsection{BHiVAE}

Now let us present our detailed model architecture. As shown in Fig \ref{fig:BHiVAE}, feed data $X$ into the encoder (parameterized as $\phi$), and in the first layer, we get the latent representation $z^1$, be divided into two parts $s^1$ and $h^1$. The part $s^1$ is the final representation part, which corresponds to feature $y^1$, and $h^1$ is the input of next layer's encoder to get latent representation $z^2$. Then through three similar network processes, we can get three representation parts $s^1,s^2,s^3$, which are disentangled, and get the part $c^3$ in the last layer, that contains information other than the above attributes of the data. All of them make up the whole representation $z = (s^1;s^2;s^3;c^3)$. The representation of each part is then mapped to the same space by a different decoder (all parameterized as $\theta$) and finally concatenated together to reconstruct the raw data, which is shown in Fig \ref{decoder}.
For the problem we discussed, we need to get the final disentangled representation $z$, i.e., we need the independence between each representation part $s^1,s^2,s^3$, and $c^3$.

\begin{figure*}[ht]
	\centering
	\subfigure[Unsupervised]{
	\label{fig:unsupervised}
	\scalebox{0.9}[0.9]{
	\includegraphics[width=0.45\linewidth]{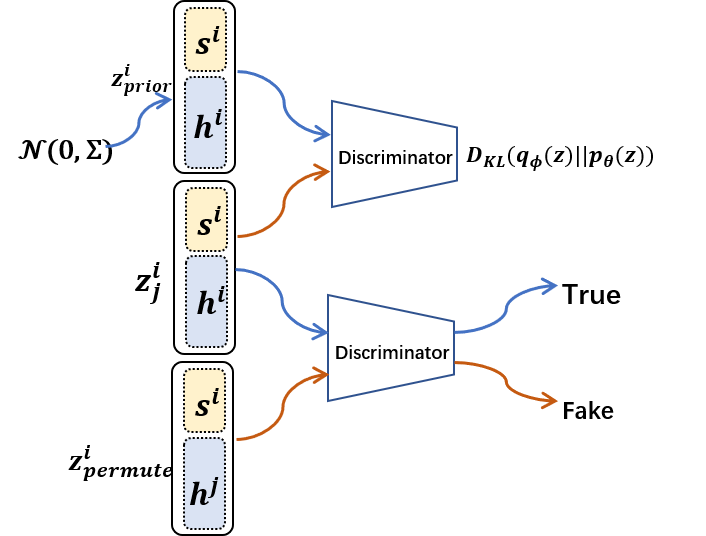}}}
	\subfigure[Supervised]{
	\label{fig:supervised}
	\scalebox{0.9}[0.9]{
	\includegraphics[width=0.4\linewidth]{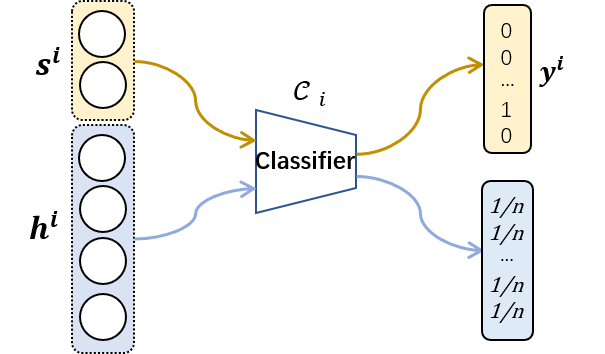}}}
	\caption{Different methods for constraining information segmentation between $s^i$ and $z^i$.}
	\label{fig:InformationSegement}
\end{figure*}

Then we can separate the whole problem into two sub-problem in $i$-th layer, so the input is $h^{i-1}$(where $h^0=x$):
\begin{itemize}
	\item[(1)] \textbf{Information flow  $h^{i-1}\rightarrow s^i\rightarrow y^i$: }Encode the upper layer's output $h^{i-1}$ to representation $z^i$, with one part $s^i$ containing sufficient information about one feature factor $y^i$;
	\item[(2)] \textbf{Information separation of $s^i$ and $h^i$:} Eliminate the information about $s^i$ in $h^i$ while requiring $s^i$ only to contain label $y^i$ information.
\end{itemize}

The first subproblem can be regarded as IB problem, the goal is to learn a representation of $s^i$, i.e. maximally expressive about feature $y^i$ while minimally informative about input data $h^{i-1}$. So it can described as:
\begin{align}
	\min I(h^{i-1};s^i)-\beta I(s^i;y^i)
	\label{sub1}
\end{align}

To satisfy the second subproblem is a complex issue, and it requires different methods to achieve it with different known conditions. So we will introduce these in follow conditions in detail. In summary, our representation is designed to enhance the internal correlation of each block while reducing the relationships between them to achieve the desired disentanglement goal.
\subsubsection{Supervised BHiVAE}

\quad In supervised case, we denote the input of $i$-th layer as $h^{i-1}$ ($h^0=x$). Given the $i$-th layer label $y^i$, we require the representation part $s^i$ to predict the feature correctly while being as compressed as possible. So the objective in $i$-th ($i=1,2,3$) layer can be described as with information measure:
\begin{align}
	 \mathcal{L}_{sup}^{class}(i)=I(h^{i-1};s^i)-\beta I(s^i;y^i)
\end{align}
We can get a upper bound of it:
\begin{align}
	\mathcal{L}_{sup}^{class}(i)&=I(h^{i-1};s^i)-\beta I(s^i;y^i)\nonumber\\
	&\le\mathbb{E}_{p(h^{i-1})}[D_{KL}(q_{\phi}(s^i|h^{i-1})\|p(s))]\nonumber\\
	&-\beta\mathbb{E}_{p(z^{i-1},y^i)}[\mathbb{E}_{q_{\phi}(s^i|h^{i-1})}[\log p_{\theta}(y^i|s^i)]]\nonumber\\
	&\triangleq\mathcal{L}_{sup}^{class_{up}}(i)
	\label{sup_class}
\end{align}
So we need one more classifier $\mathcal{C}_i$ in Fig \ref{fig:supervised} to predict $y^i$ with $s^i$.

For the second requirement, since $s^i$ is completely informative about $y^i$ which constrained in first subproblem, the elimination of information about $y^i$ is required for $h^i$:
\begin{align}
	\mathcal{L}_{info}^{sup}(i)&= I(h^i,y^i)\nonumber\\
	&=H(y^i)-H(y^i|h^i)
	\label{sup_information}
\end{align}
$H(y^i)$ is a constant, so minimizing $\mathcal{L}_{info}^{sup}(i)$ is equal to minimize:
\begin{align}
 \mathcal{L}_{info}^{sup_e}(i)= -H(y^i|h^i)
 \label{sup_e_info}
\end{align}
This is like a principle of maximum entropy, just requiring $h^i$ can't predict the factor feature $y^i$ at all, i.e. the probability predicted by $h^i$ of each category is $\frac{1}{n_i}$ ($n_i$ denotes the number of $i$-th feature categories). And $h^i$ shares the classifier $\mathcal{C}_i$ with $s^i$ as Fig \ref{fig:supervised} shows.

So in our supervised model, we can get the total objective as:
\begin{align}
	\min \{\mathcal{L}^{sup}&=\sum_{i=1}^n\mathcal{L}_{class}^{sup}(i)+\gamma\mathcal{L}_{info}^{sup_e}(i)\}
	\label{sup_hi}
\end{align}
where $\beta$ and $\gamma$ in the objective are hyper-parameter.
The objective (\ref{sup_hi}) satisfies two requirement we need, and deal with the second subproblem with a novel approach.
\vspace{-2mm}

\subsubsection{Unsupervised BHiVAE}

\quad In the unsupervised case, we know nothing about the data source, so we can only use reconstruction to constrain the representation. However, only reconstruction is not enough for disentanglement problem \cite{locatello2019challenging}, so we try to use an unique representation prior distribution to guide the representation learning. We know that all disentanglement models of the VAE series match the posterior distribution $q_{\phi}(z|x)$ to standard normal distribution prior $\mathcal{N}(0, I)$, and they can get disentanglement representation in each dimension because of the independence across $\mathcal{N}(0, I)$. For meeting our neural block representation set, we set the prior distribution $p(z)$ as $\mathcal{N}(0,\Sigma)$, where $\Sigma$ is a block diagonal symmetric matrix. Of course, the dimension of each block corresponds to the segmentation of each hidden layer. In the unsupervised model, the target is reconstruction, so we can decompose Eq. (\ref{sub1}) as:
\begin{align}
	\min &I(h^{i-1};s^i)-\beta I(s^i;x)\nonumber\\
	&\le\mathbb{E}_{p(h^{i-1})}[D_{KL}(q(z^i|h^{i-1})\|p(z))] \label{A}\\
	&-D_{KL}(q_{\phi}(z^i)\|p(z))\label{B}\\
	&-\beta[\mathbb{E}_{p(h^{i-1},y^i)}[\mathbb{E}_{q_{\phi}(s^i|h^{i-1})}[\log p_{\theta}(x|s^i)]]\label{C}\\
	&-D_{KL}(q_{\phi}(z^{i-1})\|p_{D}(x))]
	\label{D}
	\end{align}
											
The first two terms are meant to constrain the capacity of representation $z^i$, and the last two reinforce the reconstruction. VAEs model use (\ref{A}) and (\ref{C}) to achieve, and adversarial autoencoder \cite{makhzani1016adversarial} use the KL divergence (\ref{B}) between the posterior distribution $q_{\phi}(z^i)$ and prior $p(z)$ to constrain the capacity of representation and get better representation.

In our model, we also minimize the KL divergence between the posterior distribution $q_{\phi}(z^i)$ and prior $\mathcal{N}(0,\Sigma)$, i.e., $D_{KL}(q_{\phi}(z^i)\|\mathcal{N}(0,\Sigma))\rightarrow 0$. And we choose the determinstic encoder, so we get the objective:
\begin{align}
    \mathcal{L}_{recon}^{uns}=&D_{KL}(q_{\phi}(z^i)\|\mathcal{N}(0,\Sigma))\nonumber\\
    &-\beta\mathbb{E}_{p(h^{i-1})}[\mathbb{E}_{q_{\phi}(s^i|h^{i-1})}[\log p_{\theta}(x|s^i)]]
	\label{un_recon}
\end{align}
We use a discriminator at the top of Fig \ref{fig:unsupervised} to estimate and optimize $D_{KL}(p_{\phi}(h^i)\|\mathcal{N}(0,\Sigma))$.

Unlike the supervised case, we adopt a different method to satisfy the information separation requirement. When $s^i$ and $h^i$ are independent in probability, the mutual information between them comes to zero, i.e., no shared information between $s^i$ and $h^i$.
Here we apply an alternative definition of mutual information, Total Correlation (TC) penalty \cite{pmlr-v80-kim18b,watanabe1960information}, which is a popular measure of dependence for multiple random variables.

$KL(q(z)\|q(\tilde{z}))$ where $q(\tilde{z})=\prod^{d}_{j=1}q(z_j)$ is typical TC form, and in our case, we use the form $KL(p(z^i)\|p(h^i)p(s^i))=I(h^i;s^i)$. So we can get the information separation objective as:
\begin{align}
	\mathcal{L}_{info}^{uns}(i)&= I(h^i;s^i)\\
	&=KL(p(z^i)\|p(h^i)p(s^i))
	\label{uns_information}
\end{align}

In practice, $KL$ term is intractable to compute. The multiplication of marginal distributions $p(h^i)p(s^i)$ is not analytically computable, so we take a sampling approach to simulate it. After getting the a batch of representations $\{z^i_j=(s^i_j;h^i_j)\}_{j=1}^N$ in $i$-th layer, we randomly permute across the batch for $\{s^i_j\}_{j=1}^N$ and $\{h^i_j\}_{j=1}^N$ to generate sample batch under distribution $p(z^i)p(s^i)$. But direct estimating density ratio $\frac{p(z^i)}{p(h^i)p(s^i)}$ is often impossible. Thus, with random samples,  we conduct a density ratio  method \cite{nguyen2010estimating,sugiyama2012density}: use an additional classifier $D(x)$ that distinguishes between samples from the two distributions, at the bottom of Fig \ref{fig:unsupervised}:
\begin{align}
	\mathcal{L}_{info}^{uns}(i)
	&=KL(p(z^i)\|p(h^i)p(s^i))\nonumber\\
	&=TC(z^i)\nonumber\\
	&=\mathbb{E}_{q(z)}[\log\frac{p(z^i)}{p(h^i)p(s^i)}]\nonumber\\
	&\approx\mathbb{E}_{q(z)}[\log\frac{D(z^i)}{1-D(z^i)}]
	\label{TC_term}
\end{align}
In summary, the total objective under unsupervision is:
\begin{align}
	\max \{\mathcal{L}^{unsup}=\sum_{i=1}^{n}(\mathcal{L}_{recon}^{sup}+\gamma\mathcal{L}_{info}^{sup}(i))\}
	\label{uns}
\end{align}
\section{Experiments}
In this section, we present our results in quantitative and qualitative experiments. We also perform experiments comparing with $\beta$-VAE, FactorVAE, and $\beta$-TCVAE in several classic metrics. Here are datasets used in our experiments:

\textbf{MNIST \cite{lecun1998mnist}}: handwriting digital $(28\times28\times1)$ images with 60000 train samples and 10000 test samples;

\textbf{dSprites \cite{dsprites17}}: 737280 2D shapes $(64\times64\times1)$ images procedurally generated from 6 ground truth independent latent factors: shapes (heart,oval and square), x-postion (32 values), y-position (32 values), scale (6 values) and rotation (40 values);

\textbf{CelebA (cropped version) \cite{liu2015deep}}: 202599 celebrity face $(64\times64\times3)$ images with 5 landmark locations, 40 binary attributes annotations.

In the following, we perform several qualitative and quantitative experiments on these datasets and show some results comparison in both unsupervised and supervised cases. We demonstrated the ability of our model to disentangle in the unsupervised case. Besides, we also show the representation learned in the supervised case.
\begin{figure*}[ht]
	\centering
	\subfigure[Layer1 with KL=0.61]{
		\label{fig:scatter1}
		\scalebox{0.9}[0.9]{
		\includegraphics[width=0.3\linewidth]{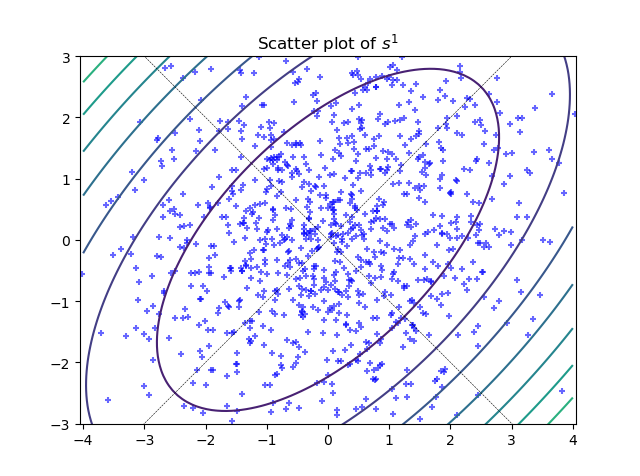}}}
	\subfigure[Layer2 with KL=0.49]{
		\label{fig:scatter2}
		\scalebox{0.9}[0.9]{
		\includegraphics[width=0.3\linewidth]{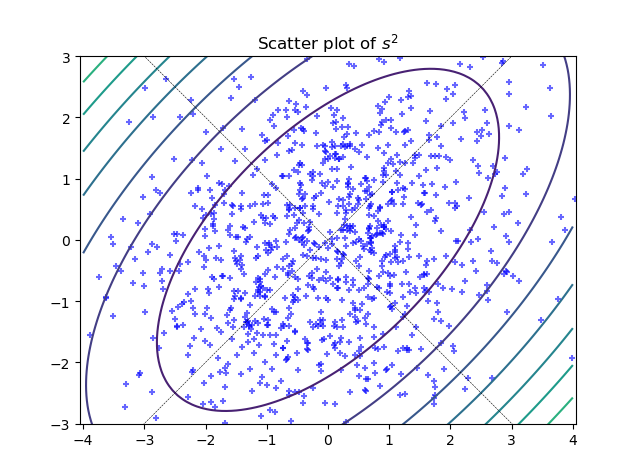}}}
	\subfigure[Layer3 with KL=0.11]{
		\label{fig:scatter3}
		\scalebox{0.9}[0.9]{
		\includegraphics[width=0.3\linewidth]{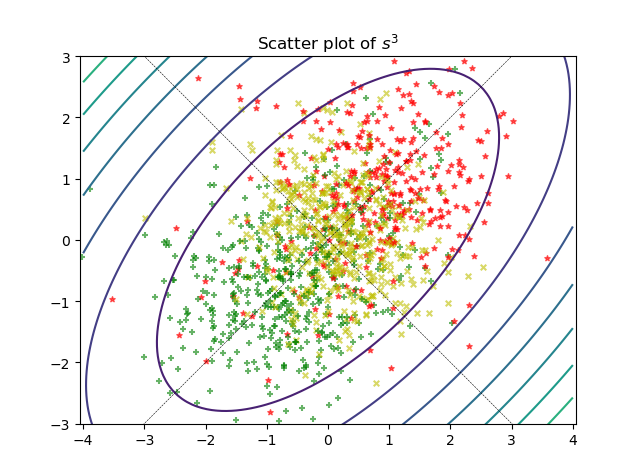}}}
	\caption{\textbf{Scatter distribution VS. Prior distribution:} Scatter plot of three layers representation $\{s^i\}_{i=1}^3$; and (C) visualizes the known category information with different colors.}
	\label{fig:scatter}
\end{figure*}

\begin{figure*}[ht]
	\centering
	\subfigure[$\beta$-VAE]{
		\label{betavaemnist}
		\scalebox{0.9}[0.9]{
		\includegraphics[width=0.22\linewidth]{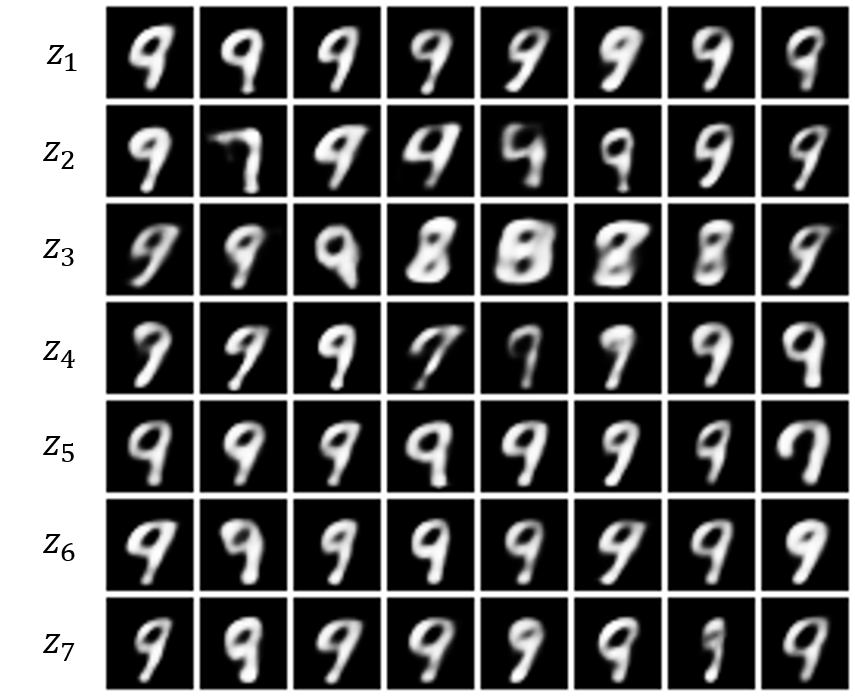}}}
	\subfigure[FactorVAE]{
		\label{factormnist}
		\scalebox{0.9}[0.9]{
		\includegraphics[width=0.20\linewidth]{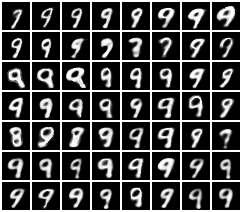}}}
	\subfigure[Guided-VAE]{
		\label{guidedmnist}
		\scalebox{0.9}[0.9]{
		\includegraphics[width=0.20\linewidth]{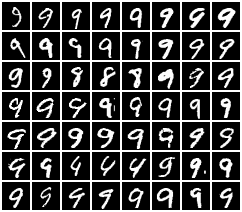}}}
	\subfigure[BHiVAE]{
		\label{himnist}
		\scalebox{0.9}[0.9]{
		\includegraphics[width=0.22\linewidth]{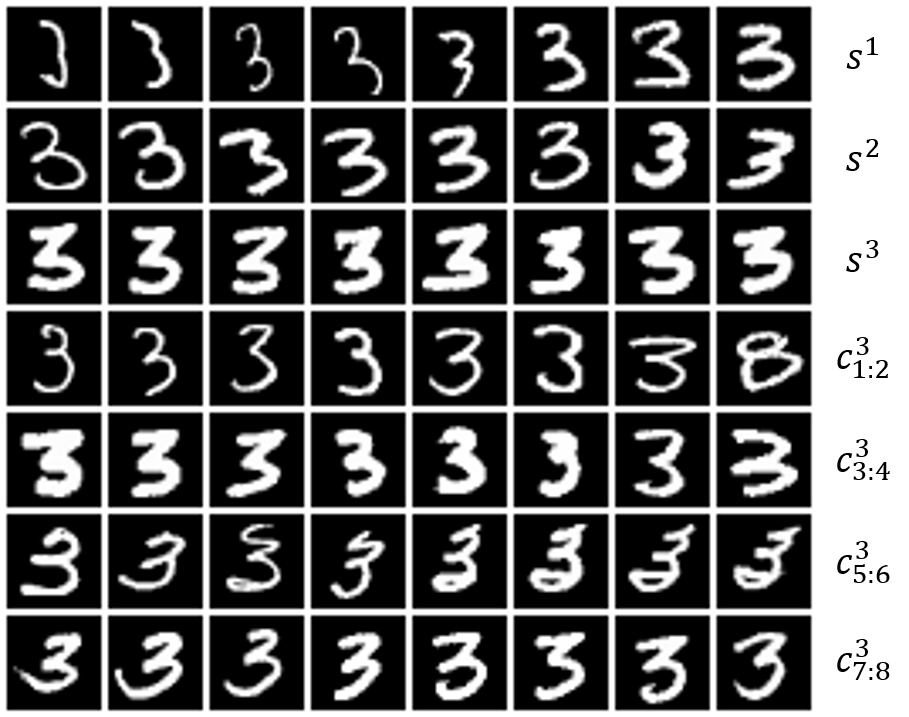}}}
	\caption{\textbf{Traversal images on MNIST:} In (a), (b) and (c), the images in $i$-th row are generated by changing $z^i$ from -3 to 3; and we change $\{s^1,s^2,s^3,c^3\}$ from (-3,-3) to (3,3), then generate the images in each row.}
	\label{fig:traversal_mnist}
\end{figure*}

\subsection{Training Details}
When training BHiVAE model, we need the encoder and decoder (Fig \ref{fig:BHiVAE}) both in supervised and unsupervised cases. 
On the CelabA dataset, we build our network with both a convolutional layer and a fully connected layer.
On the MNIST and dSprites datasets, the datasets are both $64\times64$ binary images, so we design our network to consist entirely of fully connected layers.

In evaluating the experimental results, we use the Z-differ \cite{Higgins2017betaVAELB}, SAP \cite{conf/iclr/0001SB18}, and MIG \cite{chen2018isolating} metrics to measure the quality of the disentangled representation, and observe the images generated by the traversal representation. Moreover, we use some pre-trained classifiers on attribute features to analyze the model according to the classification accuracy.

\subsection{Unsupervised BHiVAE}
In the unsupervised case, as introduced in the previous section, the most significant novel idea is we use a different prior $\mathcal{N}(0,\Sigma)$ to guide the representation learning. Additionally, we need another one to estimate the KL divergence (\ref{TC_term}). Therefore, two extra discriminators are needed for BHiVAE in Fig \ref{fig:unsupervised}. Actually, because we aim to get $D_{KL}(q_{\phi}(z^i)\|p(z))=0$, the latent representation $\{z_j^i\}_{j=1}^N$ can be considered as generated from true distribution, while prior and permuted 'presentations' $\{z^{i-perm}_j\}_{j=1}^N$ can both be considered as false. Therefore, we can simplify the network to contain only one discriminator to score these three distributions.

We want to reinforce the relationship within $s^i$ to retain the information and then decrease the dependency between $s^i$ and $h^i$ to separate information, so in our unsupervised experiments, we use this prior $\mathcal{N}(0,\Sigma)$, where

\begin{align}\scriptsize
\Sigma=\begin{bmatrix}
1&0.5&0&\cdots&0\\
0.5&1&0&\cdots&0\\
0&0&1&\cdots&0\\
\vdots&\vdots&\vdots&\ddots&\vdots\\
0&0&0&\cdots&1
\end{bmatrix}
\nonumber
\end{align}

First, we use some experiments to demonstrate the feasibility and validity of this prior setting. We train the model on the dSprites dataset first, with setting the dimension of representation $z$ to 14 ($d(z)=14$), where $d(s^i)=2,i=1,2,3$ and $d(c^3)=8$. Then we get a representation in each layer of the 1000 test images, while the three subfigures in Fig \ref{fig:scatter} shows a scatter plot of each layer representation respectively, and the curves in these figures both are the contour of the block target distribution $p(s)\sim\mathcal{N}(0,\begin{bmatrix}
1&0.5\\
0.5&1
\end{bmatrix})$.
And it is shown in Fig \ref{fig:scatter} that in the first and second layer, the distribution of $s^1$ and $s^2$ do not sufficiently match the prior $p(s)$, but as the layer forward, the KL divergence between $q_{\phi}(s^i)$ and $p(s)$ keep decresing, and the scatter plot of $s^i$ fits the prior distribution more closely. In the model, we train the encoder globally, so the front layer's representation learning can be influenced by the change of deeper representation and then yields larger KL divergence than the next layer.

Even more surprisingly, in Fig \ref{fig:scatter3}, we find that in the third layer, visualizing the 'Shape' attribute of dSprites dataset, there is an apparent clustering effect (the different colors denote different categories). This result proves our hypothesis about the deep network's ability: the deeper network is, the more detailed information it extracts. And it almost matches the prior perfectly.
Fig \ref{fig:scatter3} also gives us a better traversal way. In previous works, because only one dimension represents the attribute, they can simply change the representation from $a$ to $b$ ($a$ and $b$ both are constant). However, this does not fit our model, so the direction of the category transformation in Fig \ref{fig:scatter3} inspires us to traverse the data along the diagonal line ($y=x$). Our block prior $p(s)$ also supports that (because the prior distribution's major axis is the diagonal line too).

We perform several experiments under above architecture setting and traversal way to show the disentanglement quality on MNIST datasets. The disentanglement quantitative results of comparing with $\beta$-VAE \cite{Higgins2017betaVAELB}, FactorVAE \cite{pmlr-v80-kim18b} and Guided-VAE \cite{ding2020guided} are presented in Fig \ref{fig:traversal_mnist}. Here, considering the dimension of the representation and the number of parameters, other works' bottleneck size is set to 12, i.e., $d(z)=12$. This setting helps reduce the impact of differences in complexity between model frameworks. However, for a better comparison, we only select seven dimensions that change more regularly. In our model, we change the three-block representation $\{s^i\}_{i=1}^3$ and then the rest representation $c^3$ changes according to two dimensions as a whole, i.e., $c^3=(c^3_{1:2},c^3_{3:4},c^3_{5:6},c^3_{7:8})$.
And Fig \ref{fig:traversal_mnist} shows that $\beta$-VAE hardly ever gets a worthwhile disentangled representation, but FactorVAE appears to attribute change as representation varies. Moreover, Fig \ref{guidedmnist} and Fig \ref{himnist} both show great disentangled images, with $h_1$ changing in Guided-VAE and $s^1$ changing in BHiVAE, the handwriting is getting thicker, and $h_3,s^2$ control the angle of inclination. These all demonstrate the model capabilities of our model.
\begin{table}[h]\small
	\centering
	\scalebox{1}[1]{
		\begin{tabular}{l|ccc}
			\textbf{}          & \textbf{Z-diff $\uparrow$} & \textbf{SAP $\uparrow$} & \textbf{MIG $\uparrow$} \\ \hline
			\textbf{VAE\cite{kingma2013auto}}     & 67.1   & 0.14  & 0.23   \\
			\textbf{$\beta$-VAE\cite{Higgins2017betaVAELB}($\beta$=6)}     & 97.3   & 0.17  & 0.41   \\
			\textbf{FactorVAE\cite{pmlr-v80-kim18b}($\gamma$=7)} & 98.4      & 0.19   & 0.44   \\
			\textbf{$\beta$-TCVAE\cite{chen2018isolating}($\alpha$=1,$\beta$=8,$\gamma$=2)}   & 96.5      & 0.41  &0.49  \\
			\textbf{Guided-VAE\cite{ding2020guided}}     & \textbf{99.2}  &\textbf{0.4320} & 0.57  \\ \hline
			\textbf{BHiVAE(Ours)($\beta$=10, $\gamma$=3)}      & \textcolor{blue}{\textbf{99.0}}   & \textcolor{blue}{\textbf{0.4312}}  & \textbf{0.61}
	\end{tabular}}
	\vspace{1mm}
	\caption{\small\textbf{Disentanglement Scores:} Z-diff score, SAP score, MIG score on the dSprites dataset in the unsupervised case. The bold note the best results and blue is the second best result.}
	\label{tab1}
\end{table}

\begin{figure}[h]
	\centering
	\subfigure[$\beta$-VAE]{
		\label{betashape}
		\includegraphics[width=0.45\linewidth]{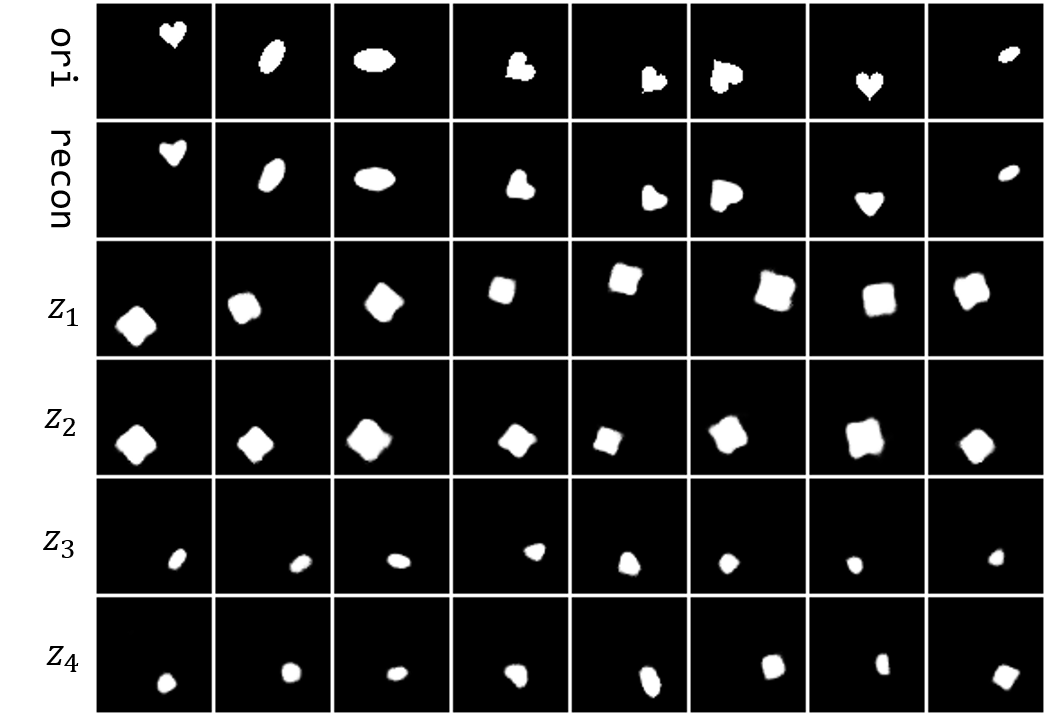}}
	\subfigure[FactorVAE]{
		\label{factorshape}
		\includegraphics[width=0.45\linewidth]{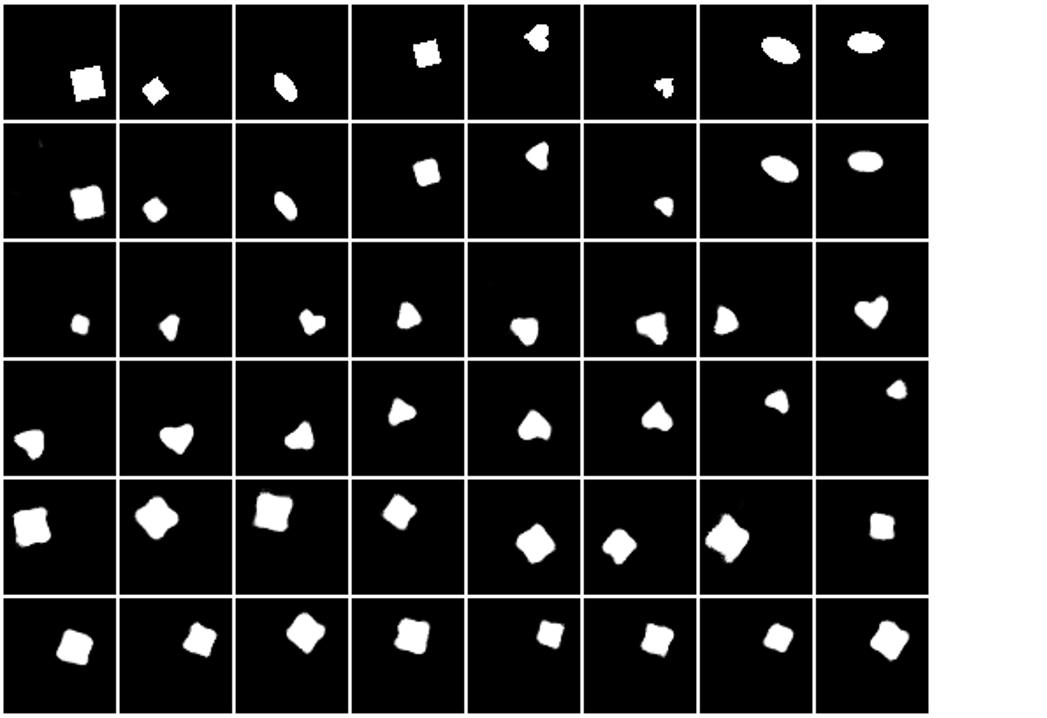}}
	
	\subfigure[GuidedVAE]{
		\label{guidedshape}
		\includegraphics[width=0.45\linewidth]{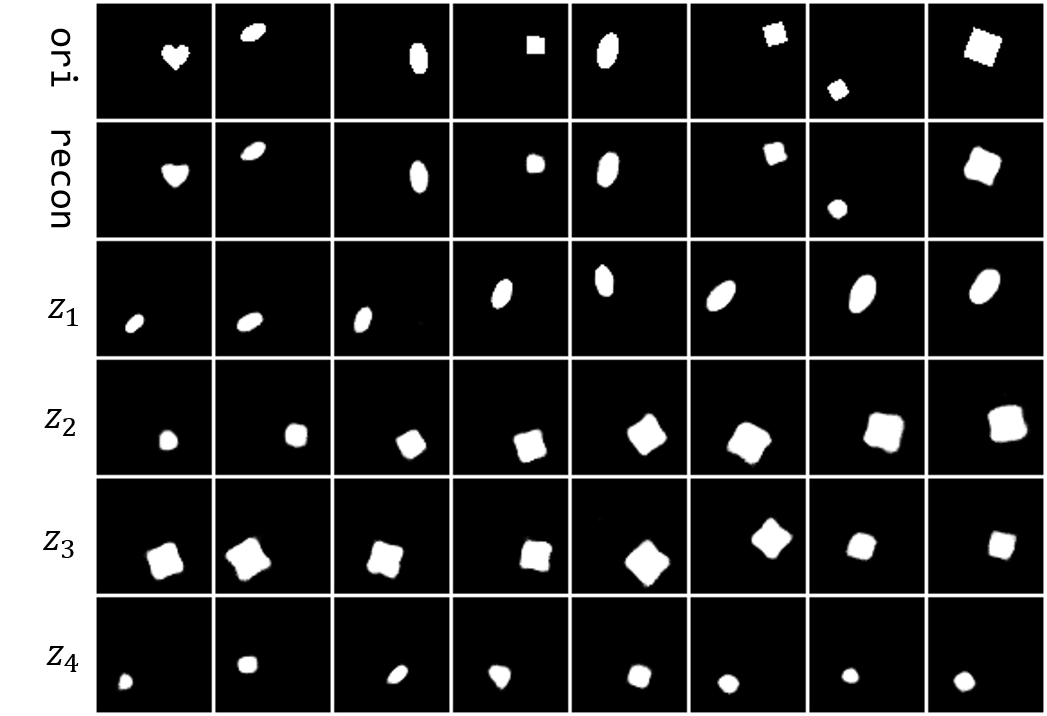}}
	\subfigure[BHiVAE(Ours)]{
		\label{hishape}
		\includegraphics[width=0.45\linewidth]{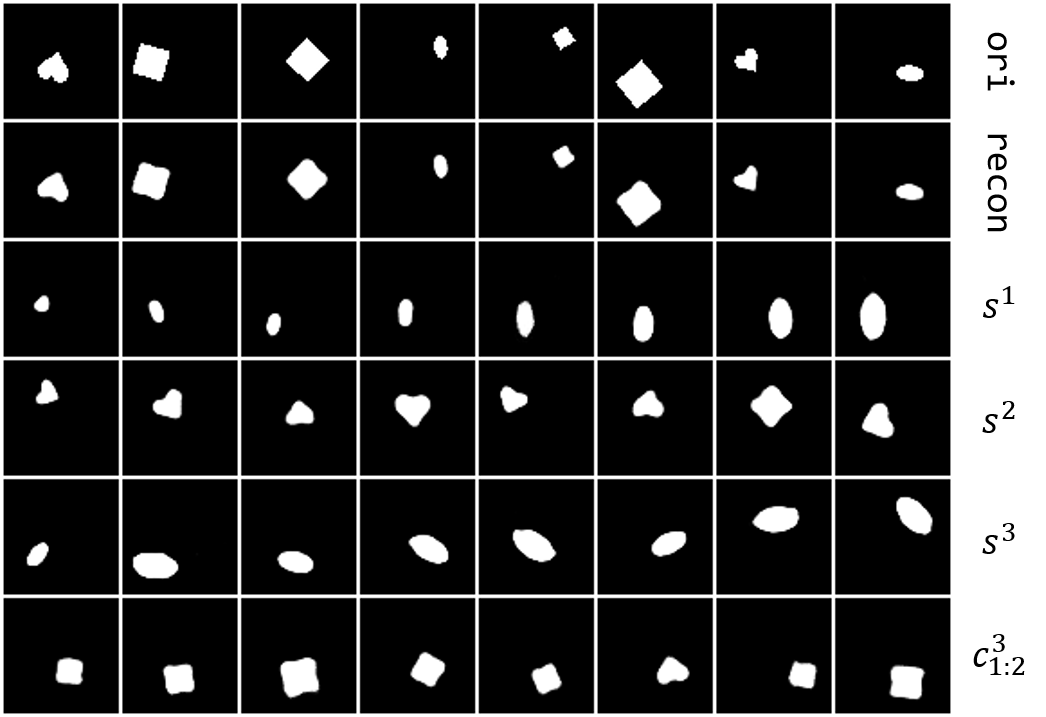}
	}
	\caption{\textbf{Traversal images on dsprites:} Images in first and second row of each figure are original and reconstruction images respectively. And others rows correspond the traversal images.}
	\label{fig:traversal_shape}
\end{figure}

We then progress to the traversal experiments on the dSprites dataset. This dataset has clear attributes distinctions, and these allow us to better observe the disentangled representation. In these experiments, BHiVAE learns a 10-dimensional representation $z=(s^1,s^2,s^3,c^3_{1:2},c^3_{3,4})$ and 8-dimensional $z=(z_1,z_2,\dots,z_8)$ in other works. We present the experiments results in Fig \ref{fig:traversal_shape} of reconstruction and traversal results. The first and second rows in four figure represent original and reconstruction images respectively. In Fig \ref{hishape}, it shows that our first three variables $s^1,s^2,s^3$ have learned the attribute characteristics (Scale, Orientation, and Position) of the data.

Moreover, we perform two quantitive experiments comparing with previous works and present our results in Table \ref{tab1} and Table \ref{tab2}. The experiments are all based on the same experiment setting in Fig \ref{fig:traversal_mnist}.

First, we compare BHiVAE with previous models with Z-differ Score \cite{Higgins2017betaVAELB}, SAP Score \cite{conf/iclr/0001SB18} and MIG Score \cite{chen2018isolating} and present the results in Table \ref{tab1}. It is clear that our model BHiVAE is at the top and that the MIG metric is better than other popular models.
The high value of the Z-diff score indicates that learned disentangled representation has less variance on the attributes of generated data as corresponding dimension changing, while SAP measures the degree of coupling between data factors and representations. Additionally MIG metric uses mutual information to measure the correlation between the data factor and learned disentangled representation, and our work is just modeled from the perspective of mutual information, which makes us performs best on the MIG score.

Not only that, but we also perform transferability experiments by conducting classification tasks on the generated representation. Here we set the representation dimensions to be the same in all models. First, we have learned a pre-trained model to obtain the representation $z$  and a pre-trained classifier to predict MNIST image label from representation. We compare the classification accuracy in Table \ref{tab2} with different dimension settings.

\begin{table}[hp!]\scriptsize
	\centering
	\scalebox{0.9}[1.05]{
		\begin{tabular}{l|ccc}
			\textbf{}          & $d_z=10 \uparrow$ & $d_z=16\uparrow$ & $d_z=32 \uparrow$ \\ \hline
			\textbf{VAE\cite{kingma2013auto}}& 97.21\%$\pm$0.42&96.62\% $\pm$ 0.51   &  96.41\%$\pm$0.22   \\
			\textbf{$\beta$-VAE\cite{Higgins2017betaVAELB}($\beta$=6)}&94.32\% $\pm$0.48    & 95.22\%$\pm$0.36   & 94.78\%$\pm$0.53   \\
			\textbf{FactorVAE\cite{pmlr-v80-kim18b}($\gamma$=7)} &93.7\%$\pm$0.07     & 94.62\%$\pm$0.12 &  93.69\%$\pm$ 0.26  \\
			\textbf{$\beta$-TCVAE\cite{chen2018isolating}($\alpha$=1,$\beta$=8,$\gamma$=2)}  & \textbf{98.4\%$\pm$0.04} &\textcolor{blue}{ \textbf{98.6\%$\pm$0.05}}  & \textcolor{blue}{\textbf{98.9\%$\pm$0.11}}   \\
			\textbf{Guided-VAE\cite{ding2020guided}}& 98.2\%$\pm$0.08   &  98.2\%$\pm$0.07&98.40\% $\pm$0.08 \\ \hline
			\textbf{BHiVAE(Ours)($\beta$=10, $\gamma$=3)}& \textcolor{blue}{\textbf{98.2\%$\pm$0.09}}& \textbf{98.7\%$\pm$0.10}  & \textbf{99.0\%$\pm$0.05}
	\end{tabular}}
	\vspace{2mm}
	\caption{\small\textbf{Accuracy of representation under unsupervised case:} The bold note the best results and blue is the second best result.}
	\label{tab2}
\end{table}

Our model appears not higher accuracy than FactorVAE and Guided-VAE in the case of $d_z=10$. That block representation setting causes a small number of factors it learns. However, as $d(z)$ is increased, our representation can learn more attribute factors of data, and then the classification accuracy can also be improved.

\subsection{Supervised BHiVAE}
In the supervised case, we still did the qualitative and quantitative experiments to evaluate our model. The same as the unsupervised case, overall autoencoder is required, and then we need a classifier to satisfy the segmentation of information at each level, as shown in Fig \ref{fig:supervised}. And we set the dimension of representation $z$ as 12 ($d(z)=12, d(c^3)=6$, and $d(s^i)=2, i=1,2,3$).

\begin{figure}[h]
	\centering
	\includegraphics[width=0.45\textwidth]{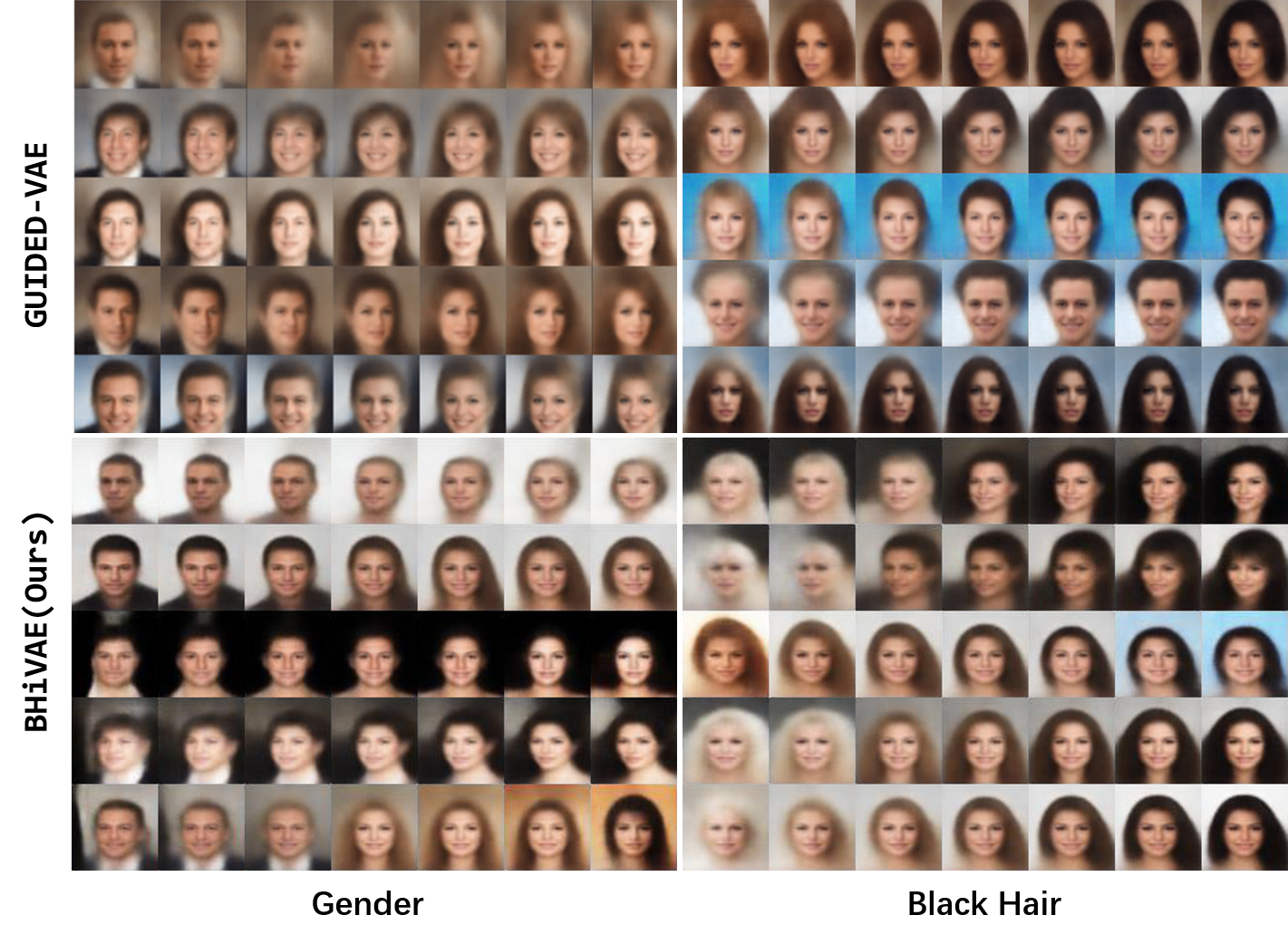}
	\caption{\textbf{Traversal Results Comparison on CelebA}: The first column is the traversal change of Gender, the second column is the change of Black Hair, the first row is from Guided-VAE \cite{ding2020guided}, the second row is ours, following the procedure of Guided-VAE.}
	\label{fig:super_celeba}
\end{figure}

We first perform several experiments comparing with Guided-VAE \cite{ding2020guided} in two attributes(Gender and Black Hair) and present the results in Fig \ref{fig:super_celeba}. When changing each attribute $s^i\in\{s^1,s^2,s^3\}$, we keep other attributes representations and content representation $c^3$ unchanged. We use the third layer representation $s^3$ to control gender attribute, while the first layers correspond to the black hair and bale, respectively.
In the supervised case, compared to Guided-VAE, we use multiple dimensions to control an attribute while Guided-VAE uses only one dimension, which may lead to insufficient information to control the traversal results. And Fig \ref{fig:super_celeba} shows that our model has a broader range of control over attributes, especially reflected in the range of hair from pure white to pure black.

\begin{figure}[htp!]
	\centering
	\scalebox{0.9}[0.9]{
	\includegraphics[width=0.45\textwidth]{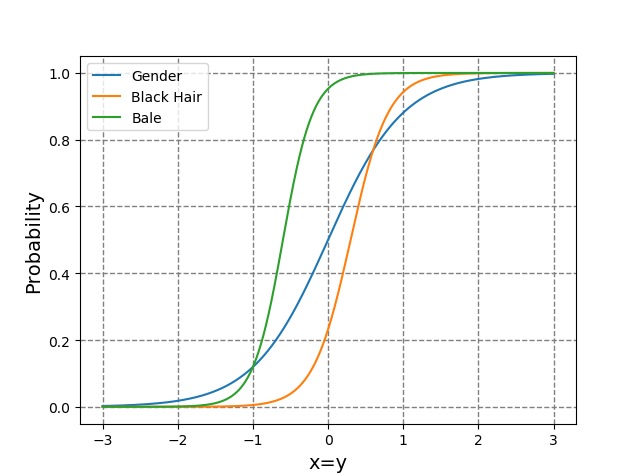}}
	\caption{\small{The classifier result used to determine if the property is available. We traverse the Black Hair ($s^1$), Bale ($s^2$) and Gender ($s^3$) attributes.}}
	\label{fig:sup_pro}
\end{figure}

Besides, our quantitative experiment is to first pre-train the BHiVAE model and three attribute classifiers of the representation and then get the representationS of the training set, traversing the three representation blocks a,b,c from $(-3,3)$ to $(3,3)$ along with the diagonal($y=x$). Fig \ref{fig:sup_pro} shows that all three attributes have a transformation threshold in the corresponding representation blocks.

\setlength{\belowcaptionskip}{-1cm} 
\begin{figure}[htp!]
	\centering
	\scalebox{0.9}[0.9]{
		\includegraphics[width=0.45\textwidth]{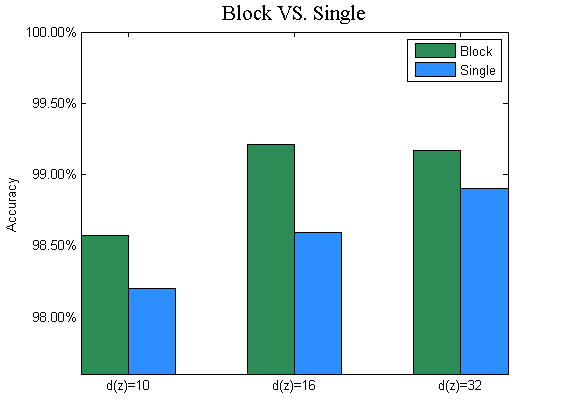}}
	\caption{\small Comparison of the accuracy of Block and Single setting model for Black Hair attribute.}
	\label{fig:block_vs_single}
\end{figure}
\subsection{Block Nodes VS. Single Node}In the previous experiments, we are all making judgments about how well the representation is disentangled and did not prove that the block setting is beneficial, so we set up the following comparison experiments for this problem.

For the comparison experiment here, we set the dimension of the model representation $z$ to 10, 16, and 32. Then in the comparison experiment, we just changed the dimension of representation $s^1$ (black hair) in the first layer to 1, and therefore the dimension of $c^3$ is changed to 5, 11, and 27 accordingly. First we pre-train these two models under the same conditions and learn a binary classifier that predicts the black hair attributes with representation $z$. It is shown in Fig \ref{fig:block_vs_single} that Block is better than Single in every dimension setting, and the accuracy of them has increased with increasing representation dimension. It could be that there is still some information about black hair in other representation parts of the model, and then the increasing dimension will allow more information about black hair to be preserved, getting better prediction accuracy.

\setlength{\belowcaptionskip}{-1cm} 
\section{Conclusion and Future Work}
We propose a new model, blocked and hierarchical variational autoencoder, for thinking about and solving disentanglement problems entirely through the perspective of information theory. We innovatively propose a blocked disentangled representation and hierarchical architecture. Then, following the idea of information segmentation, we use different methods to guide information transfer in unsupervised and supervised cases. Outstanding performance in both image traversal and representation learning allows BHiVAE to have a wider field of application.


\end{document}